# To image, or not to image: Class-specific diffractive cameras with all-optical erasure of undesired objects

Bijie Bai [†,1,2,3], Yi Luo[†,1,2,3], Tianyi Gan[1,3], Jingtian Hu[1,2,3], Yuhang Li[1,2,3], Yifan Zhao[1,3], Deniz Mengu[1,2,3], Mona Jarrahi[1,3], and Aydogan Ozcan[*,1,2,3]

[1]Electrical and Computer Engineering Department, University of California, Los Angeles, CA, 90095, USA.

[2]Bioengineering Department, University of California, Los Angeles, 90095, USA.

[3]California NanoSystems Institute (CNSI), University of California, Los Angeles, CA, USA.

[†]Equal contributing authors

[*]Correspondence: Aydogan Ozcan. Email: ozcan@ucla.edu

**Abstract**

Privacy protection is a growing concern in the digital era, with machine vision techniques widely used throughout public and private settings. Existing methods address this growing problem by, e.g., encrypting camera images or obscuring/blurring the imaged information through digital algorithms. Here, we demonstrate a camera design that performs class-specific imaging of target objects with instantaneous all-optical erasure of other classes of objects. This diffractive camera consists of transmissive surfaces structured using deep learning to perform selective imaging of target classes of objects positioned at its input field-of-view. After their fabrication, the thin diffractive layers collectively perform optical mode filtering to accurately form images of the objects that belong to a target data class or group of classes, while instantaneously erasing objects of the other data classes at the output field-of-view. Using the same framework, we also demonstrate the design of class-specific permutation cameras, where the objects of a target data class are pixel-wise permuted for all-optical class-specific encryption, while the other objects are irreversibly erased from the output image. The success of class-specific diffractive cameras was experimentally demonstrated using terahertz (THz) waves and 3D-printed diffractive layers that selectively imaged only one class of the MNIST handwritten digit dataset, all-optically erasing the other handwritten digits. This diffractive camera design can be scaled to different parts of the electromagnetic spectrum, including, e.g., the visible and infrared wavelengths, to provide transformative opportunities for privacy-preserving digital cameras and task-specific data-efficient imaging.

**Introduction**

Digital cameras and computer vision techniques are ubiquitous in modern society. Over the past few decades, computer vision-assisted applications have been adapted massively in a wide range of fields[1–3], such as video surveillance[4,5], autonomous driving assistance[6,7], medical imaging[8], facial recognition, and body motion tracking[9,10]. With the comprehensive deployment of digital cameras in workspaces and public areas, a growing concern for privacy has emerged due to the tremendous amount of image data being collected continuously[11–14]. Some commonly used methods address this concern by applying post-processing algorithms to conceal sensitive information from the acquired images[15]. Following the computer vision-aided detection of the sensitive content, traditional image redaction algorithms, such as image blurring[16,17], encryption[18,19], and image inpainting[20,21] are performed to secure private information such as human faces, plate numbers, or background objects. In recent years, deep learning techniques have further strengthened these algorithmic privacy preservation methods in terms of their robustness and speed[22–24]. Despite the success of these software-based privacy protection techniques, there exists an intrinsic risk of raw data exposure given the fact that the subsequent image processing is executed after the raw data recording/digitization and transmission, especially when the required digital processing is performed on a remote device, e.g., a cloud-based server.

Another set of solutions to such privacy concerns can be implemented at the hardware/board level, in which the data processing happens right after the digital quantization of an image, but before its transmission. Such solutions protect privacy by performing in-situ image modifications using camera-integrated online processing modules. For instance, by embedding a digital signal processor (DSP) or Trusted Platform Module (TPM) into a smart camera, the sensitive information can be encrypted or deidentified[25–27]. These camera integration solutions provide an additional layer of protection against potential attacks during the data transmission stage; however, they do not completely resolve privacy concerns as the original information is already captured digitally, and adversarial attacks can happen right after the camera's digital quantization.

Implementing these image redaction algorithms or embedded DSPs for privacy protection also creates some environmental impact as a compromise. To support the computation/processing of massive amounts of visual data being generated every day[28], i.e., billions of images and millions of hours of videos, the demand for digital computing power and data storage space rapidly increases, posing a major challenge for sustainability[29–32].

Intervening into the light propagation and image formation stage and passively enforcing privacy before the image digitization can potentially provide more desired solutions to both of these challenges outlined earlier. For example, some of the existing works use customized optics or sensor read-out circuits to modify the image formation models, so that the sensor only captures low-resolution images of the scene and, therefore, the identifying information can be concealed[33–35]. Such methods sacrifice the image quality of the entire sample field-of-view (FOV) for privacy preservation, and therefore, a delicate balance between the final image quality and privacy preservation exists; a change in this balance for different objects can jeopardize imaging performance or privacy. Furthermore, degrading the image quality of the entire FOV limits the applicable downstream tasks to low-resolution operations such as human pose estimation. In fact, sacrificing the entire image quality can be unacceptable under some circumstances such as e.g., in autonomous driving. Additionally, since these methods establish a blurred or low-resolution pixel-to-pixel mapping between the input scene and the output image, the original information of the samples can be potentially retrieved via digital inverse models, using e.g., blind image deconvolution or estimation of the inherent point-spread function.

Here, we present a new camera design using diffractive computing, which images the target types/classes of objects with high fidelity, while all-optically and instantaneously erasing other types of objects at its output (Fig. 1). This computational camera processes the optical modes that carry the sample information using successive diffractive layers optimized through deep learning by minimizing a training loss function customized for class-specific imaging. After the training phase, these diffractive layers are fabricated and assembled together in 3D, forming a computational imager between an input FOV and an output plane. This camera design is not based on a standard point-spread function, and instead the 3D-assembled diffractive layers collectively act as an optical mode filter that is statistically optimized to pass through the major modes of the target classes of objects, while filtering and scattering out the major representative modes of the other classes of objects (learned through the data-driven training process). As a result, when passing through the diffractive camera, the input objects from the target classes form clear images at the output plane, while the other classes of input objects are all-optically erased, forming non-informative patterns similar to background noise, with lower light intensity. Since all the spatial information of non-target object classes is instantaneously erased through light diffraction within a thin diffractive volume, their direct or low-resolution images are never recorded at the image plane, and this feature can be used to reduce the image storage and transmission load of the camera. Except for the illumination light, this object class-specific camera design does not utilize external computing power and is entirely based on passive transmissive layers, providing a highly power-efficient solution to task-specific and privacy-preserving imaging.

We experimentally demonstrated the success of this new class-specific camera design using THz radiation and 3D-printed diffractive layers that were assembled together (Fig. 1) to specifically and selectively image only one data class of the MNIST handwritten digit database[36], while all-optically rejecting the images of all the other handwritten digits at its output FOV. Despite the random variations observed in handwritten digits (from human to human), our analysis revealed that any arbitrary handwritten digit/class or group of digits could be selected as the target, preserving the same all-optical rejection/erasure capability for the remaining classes of handwritten digits. We also demonstrated class-specific imaging of input FOVs with multiple objects simultaneously present, where only the objects that belong to the target class were imaged at the output plane, while the rest were all-optically erased. Apart from direct imaging of the target objects from specific data classes, we further demonstrated that this diffractive imaging framework can be used to design class-specific permutation cameras that output pixel-wise permuted images of the target class of objects, while all-optically erasing other types of objects at the output FOV.

The teachings of this diffractive camera design can inspire future imaging systems that consume orders of magnitude less computing and transmission power as well as less data storage, helping with our global need for task-specific, data-efficient and privacy-aware modern imaging systems.

## Results

### Class-specific imaging using diffractive cameras

We first numerically demonstrate the class-specific camera design using the MNIST handwritten digit dataset, to selectively image handwritten digit '2' (the object class of interest) while instantaneously erasing the other handwritten digits. As illustrated in Fig. 2a, a three-layer diffractive imager with phase-only modulation layers was trained under an illumination wavelength of $\lambda$. Each diffractive layer contains 120×120 trainable transmission phase coefficients (i.e., diffractive features/neurons), each with a size of ~0.53$\lambda$. The axial distance between the input/sample plane and the first diffractive layer, between any two consecutive diffractive layers, and between the last diffractive layer and the output plane were all set to ~26.7$\lambda$. The phase modulation values of the diffractive neurons at each transmissive layer were iteratively updated using a stochastic gradient-descent-based algorithm to minimize a customized loss function, enabling object class-specific imaging. For the data class of interest, the training loss terms included the normalized mean square error (NMSE) and the negative Pearson Correlation Coefficient (PCC)[37] between the output image and the input, aiming to optimize the image fidelity at the output plane for the correct class of objects. For all the other classes of objects (to be all-optically erased), we penalized the statistical similarity between the output image and the input object (see Methods section for details). This well-balanced training loss function enabled the output images from the non-target classes of objects (i.e., the handwritten digits 0, 1, 3-9) to be all-optically erased at the output FOV, forming speckle-like background patterns with lower average intensity, whereas all the input objects of the target data class (i.e., handwritten examples of digit 2) formed high-quality images at the output plane. The resulting diffractive layers that are learned through this data-driven training process are reported in Fig. 2b, which collectively function as a spatial mode filter that is data class-specific.

After its training, we numerically tested this diffractive camera design using 10,000 MNIST test digits, which were not used during the training process. Fig. 2c reports some examples of the blind testing output of the trained diffractive imager and the corresponding input objects. These results demonstrate that the diffractive camera learned to selectively image the input objects that belong to the target data class, even if they have statistically diverse styles due to the varying nature of human handwriting. As desired, the diffractive camera generates unrecognizable noise-like patterns for the input objects from all the other data classes, all-optically erasing their information at its output plane. Stated differently, the image formation is intervened at the coherent wave propagation stage for the undesired data classes, where the characteristic optical modes that statistically represent the input objects of these non-target data classes are scattered out of the output FOV of our diffractive camera.

Importantly, this diffractive camera is *not* based on a standard point-spread function-based pixel-to-pixel mapping between the input and output FOVs, and therefore, it does not automatically result in signals within the output FOV for the transmitting input pixels that statistically overlap with the objects from the target data class. For example, the handwritten digits '3' and '8' in Fig. 2c were completely erased at the output FOV, regardless of the considerable amount of common (transmitting) pixels that they statistically share with the handwritten digit '2'. Instead of developing a spatially-invariant point-spread function, our designed diffractive camera statistically learned the characteristic optical modes possessed by different training examples, to converge as an optical mode filter, where the main modes that represent the target class of objects can pass through with minimum distortion of their relative phase and amplitude profiles, whereas the spatial information carried by the characteristic optical modes of the other data classes were scattered out. The deep learning-based optimization using the training images/examples is the key for the diffractive camera to statistically learn which optical modes must be filtered out and which group of modes needs to pass through the diffractive layers so that the output images accurately represent the spatial features of the input objects for the correct data class. As detailed in the Methods section, the training loss function and its penalty terms for the target data class and the other classes are crucial for achieving this performance.

In addition to these results summarized in Fig. 2, the same class-specific imaging system can also be adapted to selectively image input objects of other data classes by simply re-dividing the training image dataset into desired/target vs. unwanted classes of objects. To demonstrate this, we show different diffractive camera designs in Supplementary Fig. S1, where the same class-specific performance was achieved for the selective imaging of e.g., handwritten test objects from digits '5' or '7', while all-optically erasing the other data classes at the output FOV. Even more remarkable, the diffractive camera design can also be optimized to selectively image a desired *group of data classes*, while still rejecting the objects of the other data classes. For example, Supplementary Fig. S1 reports a diffractive camera that successfully imaged handwritten test objects belonging to digits '2', '5', and '7' (defining the target group of data classes), while erasing all the other handwritten digits all-optically. Stated differently, the diffractive camera was in this case optimized to selectively image three different data classes in the same design, while successfully filtering out the remaining data classes at its output FOV (see Supplementary Fig. S1).

Next, we evaluated the diffractive camera's performance with respect to the number of transmissive layers in its design (see Fig. 3 and Supplementary Fig. S1). Except for the number of diffractive layers, all the other hyperparameters of these camera designs were kept the same as before, for both the training and testing procedures. The patterns of the converged diffractive layers of each camera design are illustrated in Supplementary Fig. S2. The comparison of the class-specific imaging performance of these diffractive cameras with different numbers of trainable transmissive layers can be found in Fig. 3. Improved fidelity of the output images corresponding to the objects from the target data class can be observed as the number of diffractive layers increases, exhibiting higher image contrast, closely matching the input object features (Fig. 3a). At the same time, for the input objects from the non-target data classes, all the three diffractive camera designs generated unrecognizable noise-like patterns, all-optically erasing their information at the output. The same depth advantage can also be observed when another digit or a group of digits were selected as the target data classes. As shown in Supplementary Fig. S1, the five-layer diffractive camera designs imaged the target classes of objects with higher fidelity and contrast compared to their three-layer counterparts.

We also quantified the blind testing performance of each diffractive camera design by calculating the average PCC value between the output images and the ground truth (i.e., input objects); see Fig. 3b. For this quantitative analysis, the MNIST testing dataset was first divided into target class objects ($n_1 = 1032$ handwritten test objects for digit '2') and non-target class objects ($n_2 = 8968$ handwritten test objects for all the other digits), and the average PCC value was calculated separately for each object group. For the target data class of interest, the higher PCC value presents an improved imaging fidelity. For the other, non-target data classes, however, the *absolute* PCC values were used as an "erasure figure-of-merit", as the PCC values close to either 1 or -1 can indicate interpretable image information, which is undesirable for object erasure. Therefore, the average PCC values of the target class objects ($n_1$) and the average absolute PCC values of the non-target classes of objects ($n_2$) are presented in the first two charts in Fig. 3b. The depth advantage of the class-specific diffractive camera designs is clearly demonstrated in these results, where a deeper diffractive imager with e.g., five transmissive layers achieved (1) a better output image fidelity and a higher average PCC value for imaging the target class of objects, and (2) an improved all-optical erasure of the undesired objects (with a lower absolute PCC value) for the non-target data classes as shown in Fig. 3b.

In addition to these, a deeper diffractive camera also creates a stronger signal intensity separation between the output images of the target and non-target data classes. To quantify this signal-to-noise ratio advantage at the output FOV, we defined the average output intensity ratio ($R$) of the target to non-target data classes as:

$$R = \frac{\frac{1}{n_1}\sum_{i=1}^{n_1} \bar{O}_i^+}{\frac{1}{n_2}\sum_{i=1}^{n_2} \bar{O}_i^-} \tag{1}$$

where the numerator is the average output intensity of $n_1 = 1032$ test objects from the target data class (denoted as $O_i^+$), and the denominator is the average output intensity of $n_2 = 8968$ test objects from all the other data classes (denoted as $O_i^-$). The $R$ values of three-, four-, and five-layer diffractive camera designs were found to be 1.354, 1.464, and 1.532, respectively, as summarized in Fig. 3b. These quantitative results once again confirm that a deeper diffractive camera with more trainable layers exhibits a better performance in its class-specific imaging task and achieves an improved signal-to-noise ratio at its output.

**Simultaneous imaging of multiple objects from different data classes**

In a more general scenario, multiple objects of different classes can be presented in the same input FOV. To exemplify such an imaging scenario, the input FOV of the diffractive camera was divided into 3×3 subregions, and a random handwritten digit/object could appear in each subregion (see e.g., Fig. 4). Based on this larger FOV with multiple input objects, a three-layer and a five-layer diffractive camera were separately designed to selectively image the whole input plane, all-optically erasing all the presented objects from the non-target data classes (Fig. 4a). The design parameters of these diffractive cameras were the same as the cameras reported in the previous subsection, except that each diffractive layer was expanded from 120×120 to 300×300 diffractive pixels to accommodate the increased input FOV. During the training phase, 48,000 MNIST handwritten digits appeared randomly at each subregion, and the handwritten digit '2' was selected as our target data class to be specifically imaged. The diffractive layers of the converged camera designs are shown in Fig. 4b for the three-layer diffractive camera and in Fig. 4c for the five-layer diffractive camera.

During the blind testing phase of each of these diffractive cameras, the input test objects were randomly generated using the combinations of 10,000 MNIST test digits (not included in the training). Our imaging results reported in Fig. 4a reveal that these diffractive camera designs can selectively image the handwritten test objects from the target data class, while all-optically erasing the other objects from the remaining digits in the same FOV, regardless of which subregion they are located at. It is also demonstrated that, compared with the three-layer design, the deeper diffractive camera with five trained layers generated output images with improved fidelity and higher contrast for the target class of objects, as shown in Fig. 4a. At the same time, this deeper diffractive camera achieved stronger suppression of the objects from the non-target data classes, generating lower output intensities for these undesired objects.

**Class-specific permutation camera design**

Apart from directly imaging the objects from a target data class, a class-specific diffractive camera can also be designed to output pixel-wise permuted images of target objects, while all-optically erasing other types of objects. To demonstrate this class-specific image permutation as a form of all-optical encryption, we designed a five-layer diffractive permutation camera, which takes MNIST handwritten digits as its input and performs an all-optical permutation only on the target data class (e.g., handwritten digit '2'). The corresponding inverse permutation operation can be sequentially applied on the pixel-wise permuted output images to recover the original handwritten digits, '2'. The other handwritten digits, however, will be all-optically erased, with noise-like features appearing at the output FOV, before and after the inverse permutation operation (Fig. 5a). Stated differently, the all-optical permutation of this diffractive camera operates on a specific data class, whereas the rest of the objects from other data classes are irreversibly lost/erased at the output FOV.

To design this class-specific permutation camera, a random permutation matrix $\boldsymbol{P}$ was first generated (Fig. 5), which describes a unique one-to-one mapping of each image pixel at the input FOV to a new location/pixel at the output FOV. This randomly selected, desired permutation matrix $\boldsymbol{P}$ was applied to each input image $G$ and the resulting permuted image ($\boldsymbol{P}G$) was used as the ground truth throughout the training process of the permutation camera. The training loss function remained the same as in the previous five-layer diffractive design reported in Fig. 3a; however, instead of calculating the loss using the output and the input ($G$) images, this class-specific permutation camera design was optimized by minimizing the loss calculated using the output images and the permuted input images ($\boldsymbol{P}G$). The converged diffractive layers of this class-specific permutation camera are presented in Fig. 5b.

During the blind testing phase, the designed class-specific permutation camera was tested with 10,000 MNIST digits, never used in the training phase. As demonstrated in Fig. 5a, this permutation camera learned to selectively permute the input objects that belong to the target class (i.e., the handwritten digit '2'), generating output intensity patterns that closely resemble $\boldsymbol{P}G$. This class-specific all-optical permutation operation performed by the diffractive camera resulted in uninterpretable patterns of the target objects at the output FOV, which cannot be decoded without the knowledge of the permutation matrix, $\boldsymbol{P}$. On the other hand, for the input objects that belong to other data classes, the same permutation camera design generated noise-like, low-intensity patterns that do not match the permuted images ($\boldsymbol{P}G$). In fact, by applying the inverse permutation ($\boldsymbol{P^{-1}}$) operation on the output images of the diffractive camera, the original digits of interest from the target data class can be faithfully reconstructed, whereas all the other classes of objects ended up in noise-like patterns (see the last column of Fig. 5a), which illustrates the success of this class-specific permutation camera.

**Experimental demonstration of a class-specific diffractive camera**

We experimentally demonstrated the proof of concept of a class-specific diffractive camera by fabricating and assembling the diffractive layers using a 3D printer and testing it with a continuous wave source at $\lambda = 0.75$ mm (Fig. 6a). For these experiments, we trained a three-layer diffractive camera design using the same configuration as the system reported in Fig. 2, with the following changes: (1) the diffractive camera was "vaccinated" during its training phase against potential experimental misalignments[38], by introducing random displacements to the diffractive layers during the iterative training and optimization process (Fig. 6b, see the Methods section for details); (2) the handwritten MNIST objects were down-sampled to 15×15 pixels to form the 3D-fabricated input objects; (3) an additional image contrast-related penalty term was added to the training loss function to enhance the contrast of the output images from the target data class, which further improved the signal-to-noise ratio of the diffractive camera design. The resulting diffractive layers, including the pictures of the 3D-printed camera, are shown in Fig. 6b-c.

To blindly test the 3D-assembled diffractive camera (Fig. 6c), 12 different MNIST handwritten digits, including three digits from the target data class (digit '2') and nine digits from the other data classes were used as the input test objects of the diffractive camera. The output FOV of the diffractive camera (36×36 mm$^2$) was scanned using a THz detector forming the output images. The experimental imaging results of our 3D-printed diffractive camera are demonstrated in Fig. 7, together with the input test objects and the corresponding numerical simulation results for each input object. The experimental results show a high degree of agreement with the numerically expected results based on the optical forward model of our diffractive camera, and we observe that the test objects from the target data class were imaged well, while the other non-target test objects were completely erased at the output FOV of the camera. The success of these proof-of-concept experimental results further confirms the feasibility of our class-specific diffractive camera design.

**Discussion**

We reported a diffractive camera design that performs class-specific imaging of target objects while instantaneously erasing other objects all-optically, which might inspire energy-efficient, task-specific and secure solutions to privacy-preserving imaging. Unlike conventional privacy-preserving imaging methods that rely on post-processing of images after their digitization, our diffractive camera design enforces privacy protection by selectively erasing the information of the non-target objects during the light propagation, which reduces the risk of recording sensitive raw image data.

To make this diffractive camera design even more resilient against potential adversarial attacks, one can monitor the illumination intensity as well as the output signal intensity and accordingly trigger the camera recording only when the output signal intensity is above a certain threshold. Based on the intensity separation that is created by the class-specific imaging performance of our diffractive camera, an intensity threshold can be determined at the output image sensor to trigger image capture only when a sufficient number of photons are received, which would eliminate the recording of any digital signature corresponding to non-target objects at the input FOV. Such an intensity threshold-based recording for class-specific imaging also eliminates unnecessary storage and transmission of image data by only digitizing the target information of interest from the desired data classes.

In addition to securing the information of the undesired objects by all-optically erasing them at the output FOV, the class-specific permutation camera design reported in Fig. 5 can further perform all-optical image encryption for the desired class of objects, providing an additional layer of data security. Through the data-driven training process, the class-specific permutation camera learns to apply a randomly selected permutation operation on the target class of input objects, which can only be inverted with the knowledge of the inverse permutation operation; this class-specific permutation camera can be used to further secure the confidentiality of the images of the target data class.

Compared to the traditional digital processing-based methods, the presented diffractive camera design has the advantages of speed and resource savings since the entire non-target object erasure process is performed as the input light diffracts through a thin camera volume at the speed of light. The functionality of this diffractive camera can be enabled on demand by turning on the coherent illumination source, without the need for any additional digital computing units or an external power supply, which makes it especially beneficial for power-limited and continuously working remote systems.

It is important to emphasize that the presented diffractive camera system does not possess a traditional, spatially-invariant point-spread function. A trained diffractive camera system performs a learned, complex-valued linear transformation between the input and output fields that statistically represents the coherent imaging of the input objects from the target data class. Through the data-driven training process using examples of the input objects, this complex-valued linear transformation performed by the diffractive camera converged into an optical mode filter that, by and large, preserves the phase and amplitude distributions of the propagating modes that characteristically represent the objects of the target data class. Because of the additional penalty terms that are used to all-optically erase the undesired data classes, the same complex-valued linear transformation also acts as a modal filter, scattering out the characteristic modes that statistically represent the other types of objects that do not belong to the target data class. Therefore, each class-specific diffractive camera design results from this data-driven learning process through training examples, optimized via error backpropagation and deep learning.

Also, note that the experimental proof of concept for our diffractive camera was demonstrated using a spatially-coherent and monochromatic THz illumination source, whereas the most commonly used imaging systems in the modern digital world are designed for visible and near-infrared wavelengths, using broadband and incoherent (or partially-coherent) light. With the recent advancements in state-of-the-art nanofabrication techniques such as electron-beam lithography[39] and two-photon polymerization[40], diffractive camera designs can be scaled down to micro-scale, in proportion to the illumination wavelength in the visible spectrum, without altering their design and functionality. Furthermore, it has been demonstrated that diffractive systems can be optimized using deep learning methods to all-optically process broadband signals[41]. Therefore, nano-fabricated, compact diffractive cameras that can work in the visible and IR parts of the spectrum using partially-coherent broadband radiation from e.g., light-emitting diodes (LEDs) or an array of laser diodes would be feasible in the near future.

## Methods

**Forward-propagation model of a diffractive camera**

For a diffractive camera with $N$ diffractive layers, the forward propagation of the optical field can be modeled as a sequence of (1) free-space propagation between the $l^{th}$ and $(l+1)^{th}$ layers ($l = 0, 1, 2, \ldots, N$), and (2) the modulation of the optical field by the $l^{th}$ diffractive layer ($l = 1, 2, \ldots, N$), where the $0^{th}$ layer denotes the input/object plane and the $(N+1)^{th}$ layer denotes the output/image plane. The free-space propagation of the complex field is modeled following the angular spectrum approach[42]. The optical field $u^l(x,y)$ right after the $l^{th}$ layer after being propagated for a distance of $d$ can be written as[43]:

$$\mathbb{P}_d u^l(x,y) = \mathcal{F}^{-1}\{\mathcal{F}\{u^l(x,y)\}H(f_x, f_y; d)\} \tag{2}$$

where $\mathbb{P}_d$ represents the free-space propagation operator, $\mathcal{F}$ and $\mathcal{F}^{-1}$ are the two-dimensional Fourier transform and the inverse Fourier transform operations, and $H(f_x, f_y; d)$ is the transfer function of free space:

$$H(f_x, f_y; d) = \begin{cases} \exp\left\{jkd\sqrt{1 - \left(\frac{2\pi f_x}{k}\right)^2 - \left(\frac{2\pi f_y}{k}\right)^2}\right\}, & f_x^2 + f_y^2 < \frac{1}{\lambda^2} \\ 0, & f_x^2 + f_y^2 \geq \frac{1}{\lambda^2} \end{cases} \tag{3}$$

where $j = \sqrt{-1}$, $k = \frac{2\pi}{\lambda}$ and $\lambda$ is the wavelength of the illumination light. $f_x$ and $f_y$ are the spatial frequencies along the $x$ and $y$ directions, respectively.

We consider only the phase modulation of the transmitted field at each layer, where the transmittance coefficient $t^l$ of the $l^{th}$ diffractive layer can be written as:

$$t^l(x,y) = \exp\{j\phi^l(x,y)\} \tag{4}$$

where $\phi^l(x,y)$ denotes the phase modulation of the trainable diffractive neuron located at $(x,y)$ position of the $l^{th}$ diffractive layer. Based on these definitions, the complex optical field at the output plane of a diffractive camera can be expressed as:

$$o(x,y) = \mathbb{P}_{d_{N,N+1}} \left( \prod_{l=N}^{1} t^l(x,y) \cdot \mathbb{P}_{d_{l-1,l}} \right) g(x,y) \tag{5}$$

where $d_{l-1,l}$ represents the axial distance between the $(l-1)^{th}$ and the $l^{th}$ layers, $g(x,y)$ is the input optical field, which is the amplitude of the input objects (handwritten digits) used in this work.

**Training loss function**

The reported diffractive camera systems were optimized by minimizing the loss functions that were calculated using the intensities of the input and output images. The input and output intensities $G$ and $O$, respectively, can be written as:

$$G(x,y) = |g(x,y)|^2 \qquad (6)$$

$$O(x,y) = |o(x,y)|^2 \qquad (7)$$

The loss function, calculated using a batch of training input objects $\boldsymbol{G}$ with the corresponding output images $\boldsymbol{O}$ can be defined as:

$$Loss(\boldsymbol{O},\boldsymbol{G}) = Loss_+(\boldsymbol{O^+},\boldsymbol{G^+}) + Loss_-(\boldsymbol{O^-},\boldsymbol{G^-},G_k^+) \qquad (8)$$

where $\boldsymbol{O^+}, \boldsymbol{G^+}$ represent the output and input images from the target data class (i.e., desired object class), and $\boldsymbol{O^-}, \boldsymbol{G^-}$ represent the output and input images from the other data classes (to be all-optically erased), respectively.

The $Loss_+$ is designed to reduce the NMSE and enhance the correlation between any target class input object $O^+$ and its output image $G^+$, i.e.,

$$Loss_+(O^+,G^+) = \alpha_1 \times \text{NMSE}(O^+, G^+) + \alpha_2 \times \bigl(1 - \text{PCC}(O^+,G^+)\bigr) \qquad (9)$$

where $\alpha_1$ and $\alpha_2$ are constants and NMSE is defined as:

$$\text{NMSE}(O^+,G^+) = \frac{1}{MN}\sum_{m,n}\left(\frac{O^+_{m,n}}{\max(O^+)} - G^+_{m,n}\right)^2 \qquad (10)$$

$m$ and $n$ are the pixel indices of the images, and $MN$ represents the total number of pixels in each image. The output image $O^+$ was normalized by its maximum pixel value, $\max(O^+)$. The PCC value between any two images $A$ and $B$ is calculated using[37]:

$$\text{PCC}(A, B) = \frac{\sum(A - \bar{A})(B - \bar{B})}{\sqrt{\sum(A - \bar{A})^2 \sum(B - \bar{B})^2}} \qquad (11)$$

The term $\bigl(1 - \text{PCC}(O^+,G^+)\bigr)$ was used in $Loss_+$ in order to maximize the correlation between $O^+$ and $G^+$, as well as to ensure a non-negative loss value since the PCC value of any two images is always between -1 and 1.

Different from $Loss_+$, the $Loss_-$ function is designed to *reduce* (1) the *absolute* correlation between the output $O^-$ and its corresponding input $G^-$, (2) the *absolute* correlation between $O^-$ and an arbitrary object $G_k^+$ from the target class, and (3) the correlation between $O^-$ and itself shifted by a few pixels $O_{\text{sft}}^-$, which can be formulated as:

$$Loss_-(O^-, G^-, G_k^+) = \beta_1 \times |\text{PCC}(O^-, G^-)| + \beta_2 \times |\text{PCC}(O^-, G_k^+)| + \beta_3 \times \text{PCC}(O^-, O_{\text{sft}}^-) \quad (12)$$

where $\beta_1$, $\beta_2$ and $\beta_3$ are constants. Here the $G_k^+$ refers to an image of an object from the target data class in the training set, which was randomly selected for every training batch, and the subscript $k$ refers to a random index. In other words, within each training batch, the $\text{PCC}(O^-, G_k^+)$ was calculated using the output image from the non-target data class and a random ground truth image from the target class. By adding such a loss term, we prevent the diffractive camera from converging to a solution where all the output images look like the target object. The $O_{\text{sft}}^-$ was obtained using:

$$O_{\text{sft}}^-(x, y) = O^-(x - s_x, y - s_y) \quad (13)$$

where $s_x = s_y = 5$ denote the number of pixels that $O^-$ is shifted in each direction. By minimizing $\text{PCC}(O^-, O_{\text{sft}}^-)$, we forced the diffractive camera to generate uninterpretable noise-like output patterns for input objects that do not belong to the target data class.

The coefficients $(\alpha_1, \alpha_2, \beta_1, \beta_2, \beta_3)$ in the two loss functions were empirically set to (1, 3, 6, 3, 2).

**Digital implementation and training scheme**

The diffractive camera models reported in this work were trained with the standard MNIST handwritten digit dataset under $\lambda = 0.75$ mm illumination. Each diffractive layer has a pixel/neuron size of 0.4 mm, which only modulates the phase of the transmitted optical field. The axial distance between the input plane and the first diffractive layer, the distances between any two successive diffractive layers, and the distance between the last diffractive layer and the output plane are set to 20 mm, i.e., $d_{l-1,l} = 20$ mm ($l = 1, 2, \ldots, N + 1$). For the diffractive camera models that take a single MNIST image as its input (e.g., reported in Figs. 2-3), each diffractive layer contains 120×120 diffractive pixels. During the training, each 28×28 MNIST raw image was first linearly upscaled to 90×90 pixels. Next, the upscaled training dataset was augmented with random image transformations, including a random rotation by an angle within $[-10°, +10°]$, a random scaling by a factor within $[0.9, 1.1]$, and a random shift in each lateral direction by an amount of $[-2.13\lambda, +2.13\lambda]$.

For the diffractive camera model reported in Fig. 4 that takes multiplexed objects as its input, each diffractive layer contains 300×300 diffractive pixels. The MNIST training digits were first upscaled to 90×90 pixels and then randomly transformed with $[-10°, +10°]$ angular rotation, $[0.9, 1.1]$ scaling, and $[-2.13\lambda, +2.13\lambda]$ translation. Nine different handwritten digits were randomly selected and arranged into 3×3 grids, generating a multiplexed input image with 270×270 pixels for the diffractive camera training.

For the diffractive permutation camera reported in Fig. 5, each diffractive layer contains 120×120 diffractive pixels. The design parameters of this class-specific permutation camera were kept the same as the five-layer diffractive camera reported in Fig. 3a, except that the handwritten digits were down-sampled to 15×15 pixels considering that the required computational training resources for the permutation operation increase quadratically with the total number of input image pixels. The MNIST training digits were augmented using the same random transformations as described above. The 2D permutation matrix $\boldsymbol{P}$ was generated by randomly shuffling the rows of a 225×225 identity matrix. The inverse of $\boldsymbol{P}$ was obtained by using the transpose operation, i.e., $\boldsymbol{P}^{-1} = \boldsymbol{P}^{T}$. The training loss terms for the class-specific permutation camera remained the same as described in Equations (8), (9), and (12), except that the permuted input images ($\boldsymbol{P}G$) were used as the ground truth, i.e.,

$$Loss_{\text{Permutation}}(O, \boldsymbol{P}G) = Loss_{+}(O^{+}, \boldsymbol{P}G^{+}) + Loss_{-}(O^{-}, \boldsymbol{P}G^{-}, \boldsymbol{P}G_{k}^{+}) \tag{14}$$

The MNIST handwritten digit dataset was divided into training, validation, and testing datasets without any overlap, with each set containing 48,000, 12,000, and 10,000 images, respectively. The diffractive camera models reported in this paper were trained using the Adam optimizer[44] with a learning rate of 0.03. The batch size used for all the trainings was 60. All models were trained and tested using PyTorch 1.11 with a GeForce RTX 3090 graphical processing unit (NVIDIA Inc.). The typical training time for a three-layer diffractive camera (e.g., in Fig. 2) is ~21 hours for 1000 epochs.

**Experimental Design**

For the experimentally validated diffractive camera design shown in Fig. 6, an additional contrast loss $L_c$ was added to $Loss_{+}$ i.e.,

$$Loss_{+}(O^{+}, G^{+}) = \alpha_1 \times \text{NMSE}(O^{+}, G^{+}) + \alpha_2 \times (1 - \text{PCC}(O^{+}, G^{+})) + \alpha_3 \times L_c(O^{+}, G^{+}) \tag{15}$$

The coefficients $(\alpha_1, \alpha_2, \alpha_3)$ were empirically set to (1, 3, 5) and $L_c$ is defined as:

$$L_c(O^+, G^+) = \frac{\sum\left(O^+ \cdot \left(1 - \widehat{G^+}\right)\right)}{\sum(O^+ \cdot \widehat{G^+}) + \varepsilon} \tag{16}$$

where $\varepsilon = 1\mathrm{e}^{-6}$ was added to the denominator to avoid divide-by-zero error. $\widehat{G^+}$ is a binary mask indicating the transmissive regions of the input object $G^+$, which is defined as:

$$\widehat{G^+}(m,n) = \begin{cases} 1, & G^+(m,n) > 0.5 \\ 0, & otherwise \end{cases} \tag{17}$$

By adding this image contrast related training loss term, the output images of the target objects exhibit enhanced contrast which is especially helpful in non-ideal experimental conditions.

In addition, the MNIST training images were first linearly downsampled to 15×15 pixels and then upscaled to 90×90 pixels using nearest-neighbor interpolation. Then, the resulting input objects were augmented using the same parameters as described before and were fed into the diffractive camera for training. Each diffractive layer had 120×120 trainable diffractive neurons.

To overcome the challenges posed by the fabrication inaccuracies and mechanical misalignments during the experimental validation of the diffractive camera, we vaccinated our diffractive model during the training by deliberately introducing random displacements to the diffractive layers[38]. During the training process, a 3D displacement $\boldsymbol{D} = (D_x, D_y, D_z)$ was randomly added to each diffractive layer following the uniform (**U**) random distribution:

$$D_x \sim \mathbf{U}(-\Delta_{x,tr}, \Delta_{x,tr}) \tag{18}$$

$$D_y \sim \mathbf{U}(-\Delta_{y,tr}, \Delta_{y,tr}) \tag{19}$$

$$D_z \sim \mathbf{U}(-\Delta_{z,tr}, \Delta_{z,tr}) \tag{20}$$

where $D_x$ and $D_y$ denote the random lateral displacement of a diffractive layer in $x$ and $y$ directions, respectively. $D_z$ denotes the random displacement added to the axial distances between any two consecutive diffractive layers. $\Delta_{*,tr}$ represents the maximum amount of shift allowed along the corresponding axis, which was set as $\Delta_{x,tr} = \Delta_{y,tr} = 0.4$ mm (~0.53$\lambda$), and $\Delta_{z,tr} = 1.5$ mm (2$\lambda$) throughout the training process. $D_x$, $D_y$, and $D_z$ of each diffractive layer were independently sampled from the given uniform random distributions. The diffractive camera model used for the experimental validation was trained for 50 epochs.

**Experimental THz imaging setup**

We validated the fabricated diffractive camera design using a THz continuous wave scanning system. The phase values of the diffractive layers were first converted into height maps using the refractive index of the 3D printer material. Then, the layers were printed using a 3D printer (Pr 110, CADworks3D). A layer holder that sets the positions of the input plane, output plane, and each diffractive layer was also 3D printed (Objet30 Pro, Stratasys) and assembled with the printed layers. The test objects were 3D printed (Objet30 Pro, Stratasys) and coated with aluminum foil to define the transmission areas.

The experimental setup is illustrated in Fig. 5a. The THz source used in the experiment was a WR2.2 modular amplifier/multiplier chain (AMC) with a compatible diagonal horn antenna (Virginia Diode Inc.). The input of AMC was a 10 dBm RF input signal at 11.1111 GHz ($f_{RF1}$) and after being multiplied 36 times, the output radiation was at 0.4 THz. The AMC was also modulated with a 1 kHz square wave for lock-in detection. The output plane of the diffractive camera was scanned with a 1 mm step size using a single-pixel Mixer/AMC (Virginia Diode Inc.) detector mounted on an XY positioning stage that was built by combining two linear motorized stages (Thorlabs NRT100). A 10 dBm RF signal at 11.083 GHz ($f_{RF2}$) was sent to the detector as a local oscillator to down-convert the signal to 1 GHz. The down-converted signal was amplified by a low-noise amplifier (Mini-Circuits ZRL-1150-LN+) and filtered by a 1 GHz (+/-10 MHz) bandpass filter (KL Electronics 3C40-1000/T10-O/O). Then the signal passed through a tunable attenuator (HP 8495B) for linear calibration and a low-noise power detector (Mini-Circuits ZX47-60) for absolute power detection. The detector output was measured by a lock-in amplifier (Stanford Research SR830) with the 1 kHz square wave used as the reference signal. Then the lock-in amplifier readings were calibrated into linear scale. A digital 2×2 binning was applied to each measurement of the intensity field to match the training feature size used in the design phase.

**Author contributions:**

A.O. conceived the research and initiated the project. B.B., Y.L., and D.M. developed the numerical simulation codes. B.B., Y.L., T.G., J.H., Y.L., and Y.Z. performed the fabrication of the diffractive system and conducted the experiments. All the authors participated in the analysis and discussion of the results. B.B., Y.L., T.G., J.H., and A.O. prepared the manuscript and all authors contributed to the manuscript. A.O. and M.J. supervised the project.

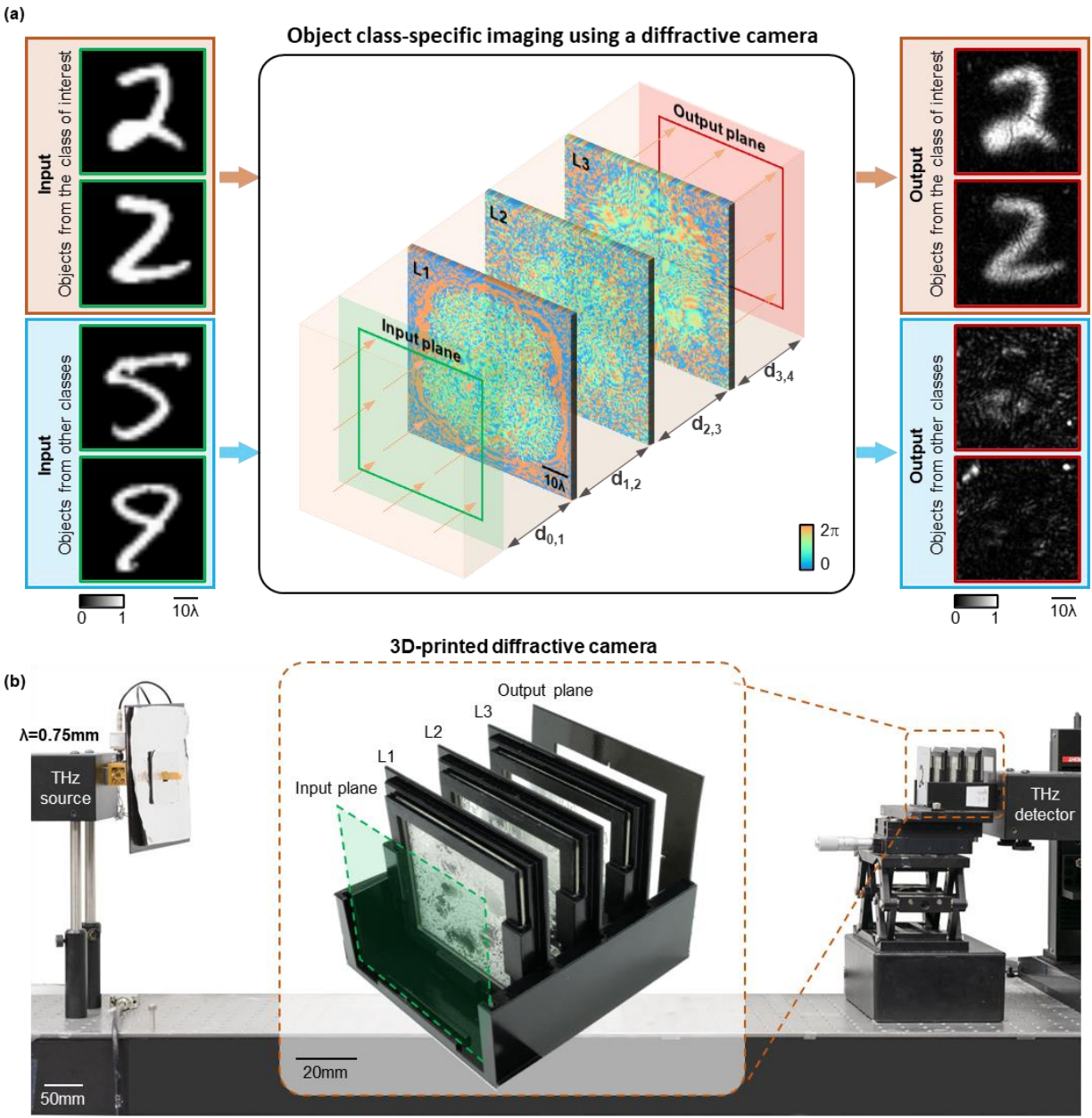

**Figure 1. Object class-specific imaging using a diffractive camera**. (a) Illustration of a three-layer diffractive camera trained to perform object class-specific imaging with instantaneous all-optical erasure of the other classes of objects at its output FOV. (b) The experimental setup for the diffractive camera testing using coherent THz illumination.

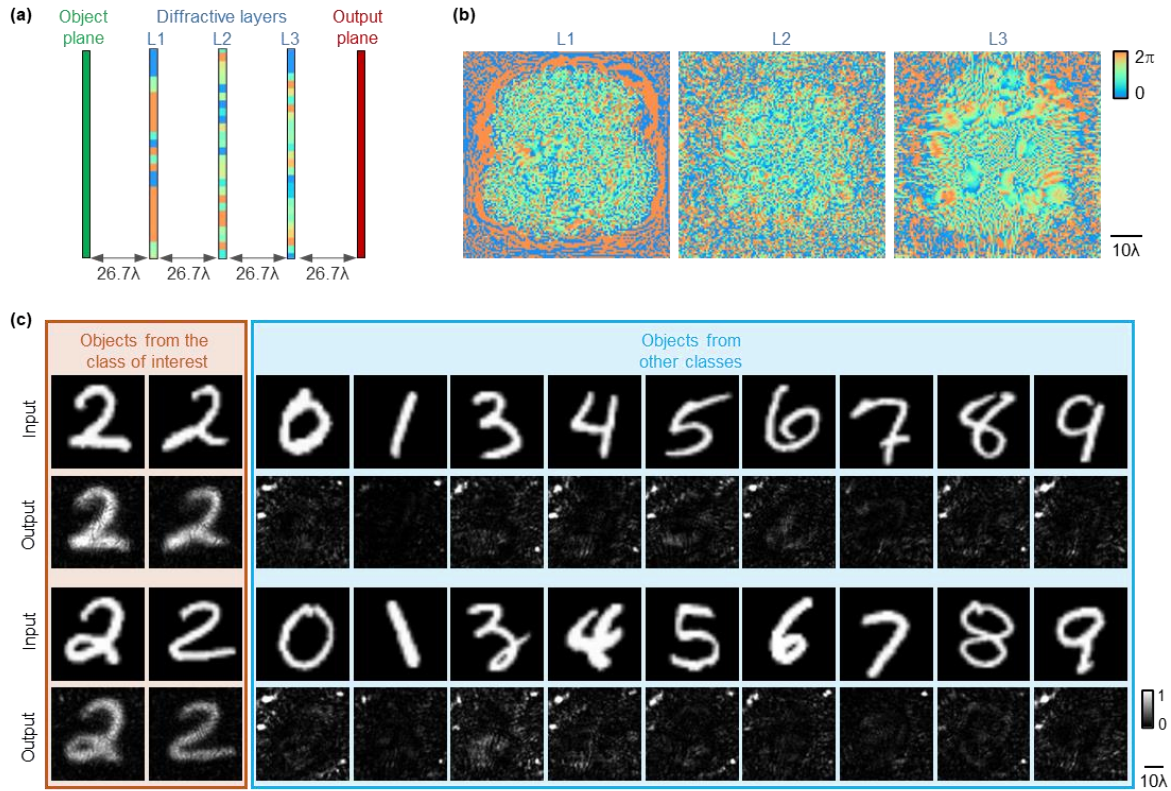

**Figure 2. Design schematic and blind testing results of the class-specific diffractive camera.** (a) The physical layout of the three-layer diffractive camera design. (b) Phase modulation patterns of the converged diffractive layers of the camera. (c) The blind testing results of the diffractive camera. The output images were normalized using the same constant for visualization.

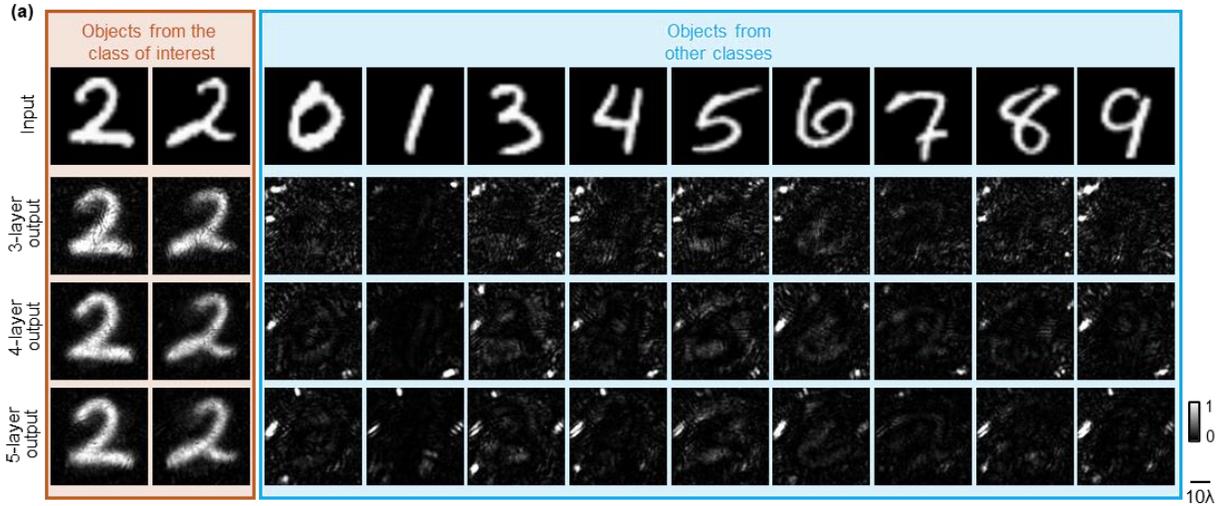

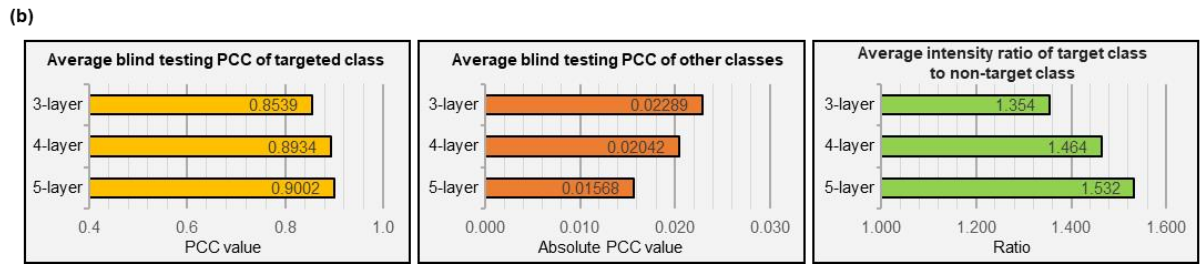

**Figure 3. Performance advantages of deeper diffractive cameras.** (a) Comparison of the output images using diffractive camera designs with three, four, and five layers. The output images at each row were normalized using the same constant for visualization. (b) Quantitative comparison of the three diffractive camera designs. The left panel compares the average PCC values calculated using input objects from the target data class only (i.e., 1032 different handwritten digits). The middle panel compares the average absolute PCC values calculated using input objects from the other data classes (i.e., 8968 different handwritten digits). The right panel plots the average output intensity ratio ($R$) of the target to non-target data classes.

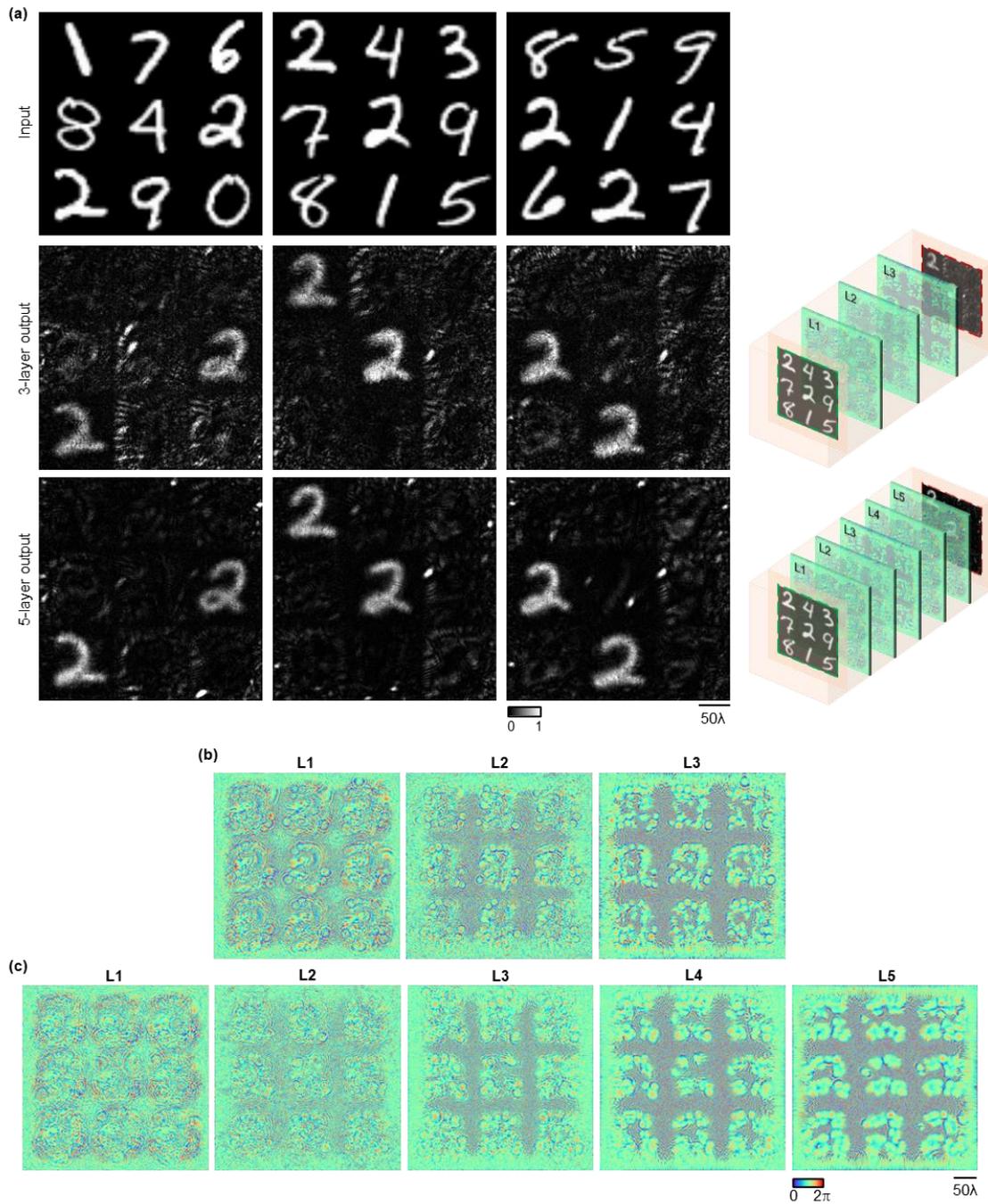

**Figure 4. Simultaneous imaging of multiple objects of different data classes using a diffractive camera**. (a) Schematic and the blind testing results of a three-layer diffractive camera and a five-layer diffractive camera. The output images in each row were normalized using the same constant for visualization. (b) Phase modulation patterns of the converged diffractive layers for the three-layer diffractive camera design. (c) Phase modulation patterns of the converged diffractive layers for the five-layer diffractive camera design.

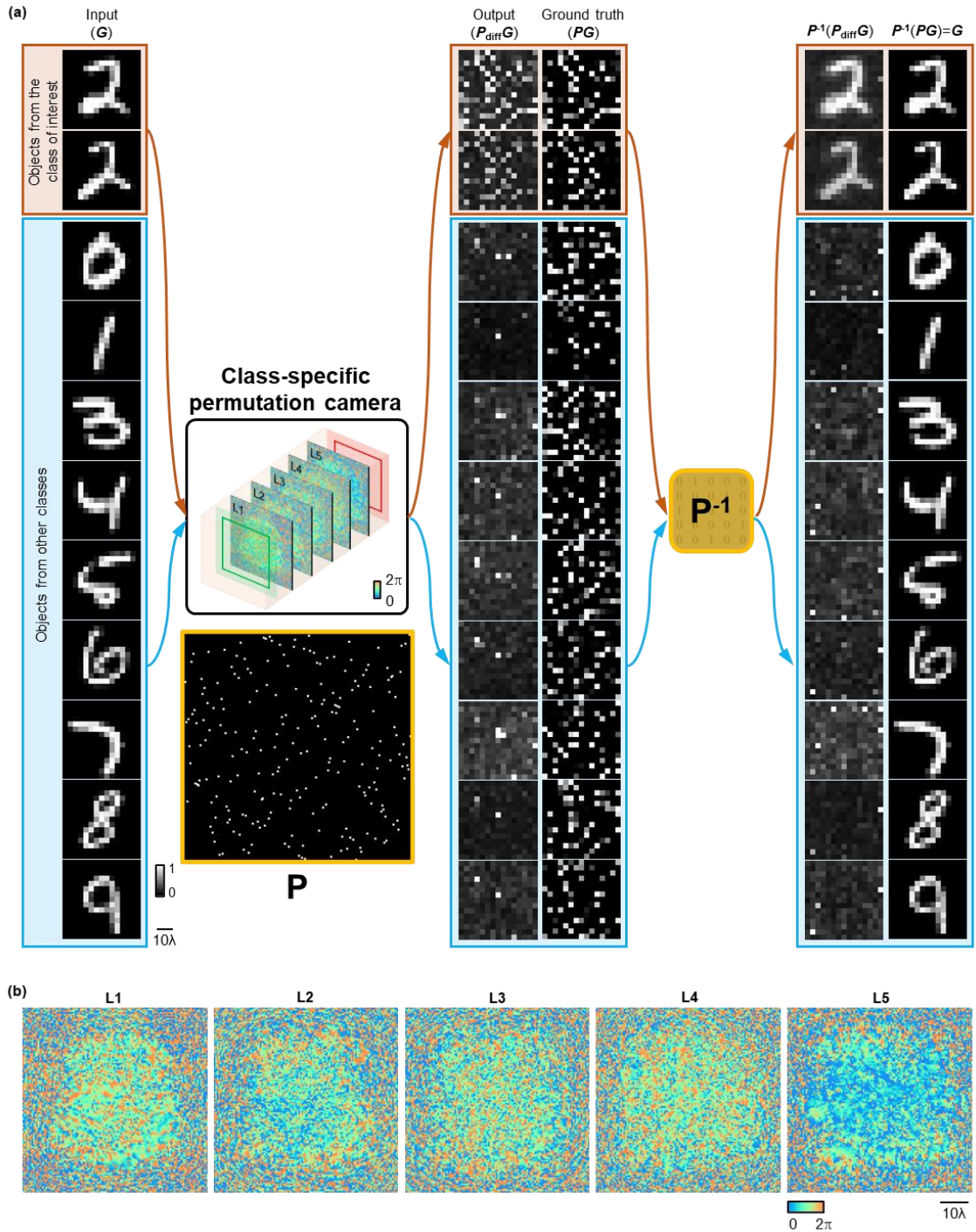

**Figure 5. Class-specific permutation camera.** (a) Illustration of a five-layer diffractive camera trained to perform class-specific permutation operation (denoted as $\boldsymbol{P}$) with instantaneous all-optical erasure of the other classes of objects at its output FOV. Application of an inverse permutation ($\boldsymbol{P^{-1}}$) to the output images recovers the original objects of the target data class, whereas the rest of the objects from other data classes are irreversibly lost/erased at the output FOV. The output images were normalized using the same constant for visualization. (b) Phase modulation patterns of the converged diffractive layers of the class-specific permutation camera.

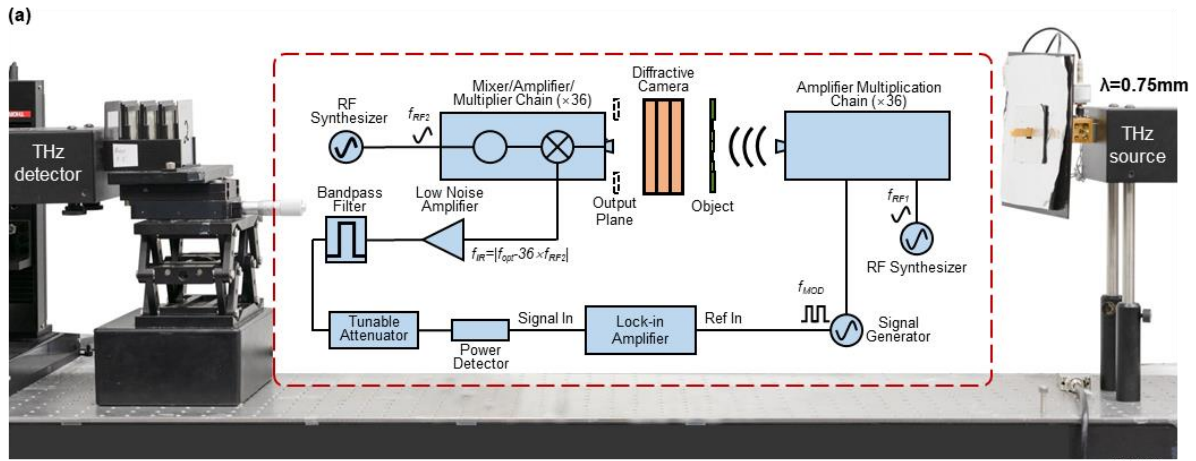

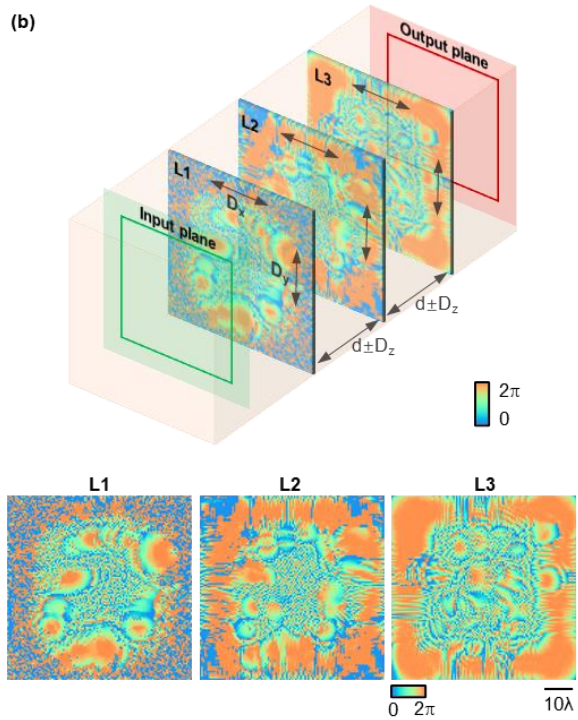

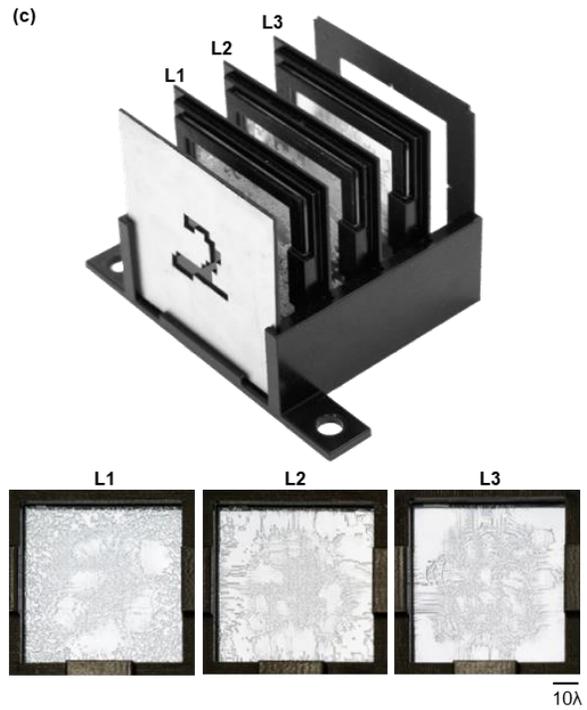

**Figure 6. Experimental setup for object class-specific imaging using a diffractive camera.** (a) Schematic of the experimental setup using THz illumination. (b) Schematic of the misalignment resilient training of the diffractive camera and the converged phase patterns. (c) Photographs of the 3D printed and assembled diffractive system.

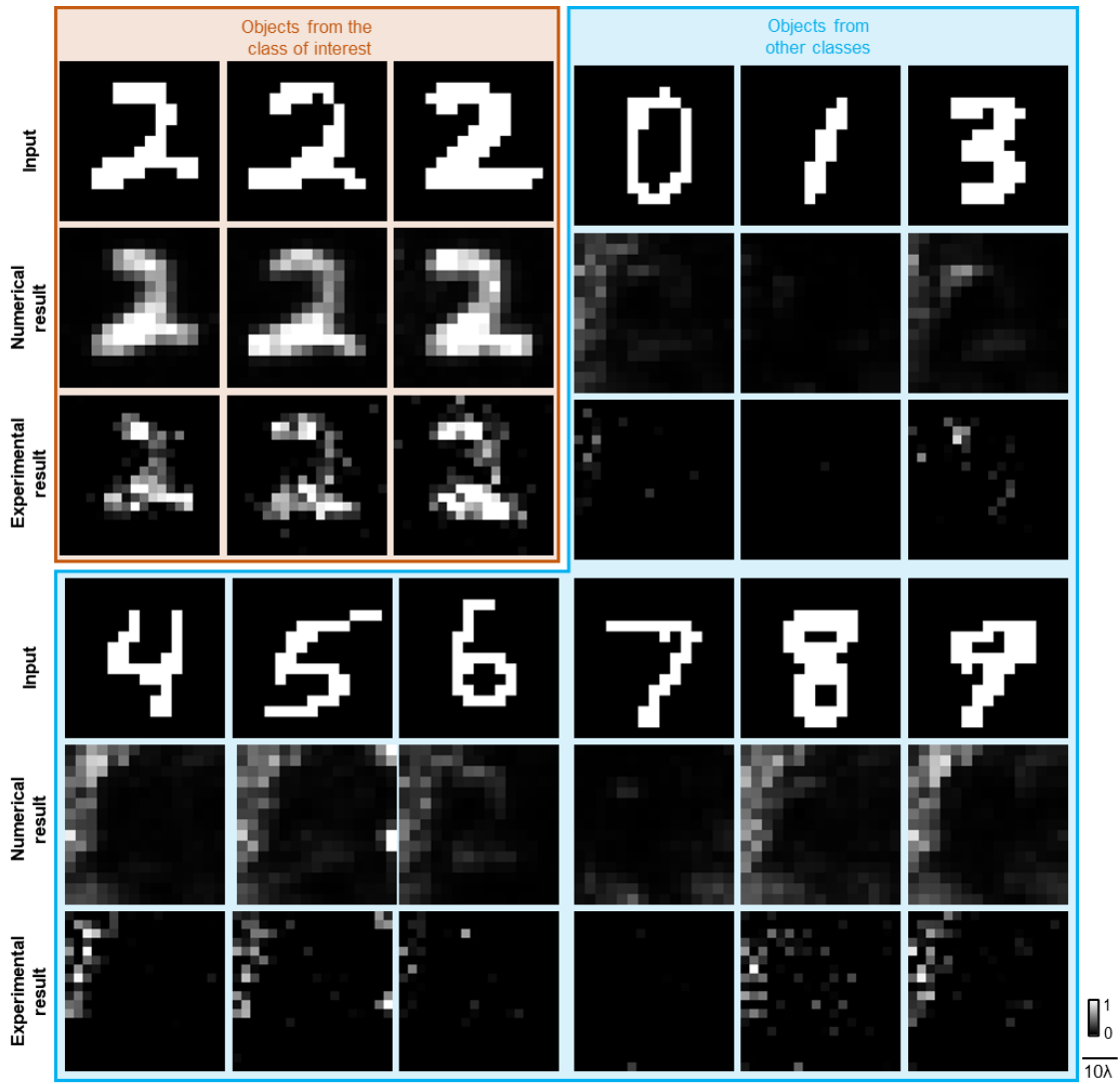

**Figure 7. Experimental results of object class-specific imaging using a 3D-printed diffractive camera.**